\documentclass[12pt,preprint]{aastex}

\newcommand{\source}{4U 1820--30}

\shorttitle{Spectral variability study of 4U 1820--30}
\shortauthors{Tarana et al.}

\begin{document}

\title{{\it INTEGRAL}{$^\star$} spectral variability study of the atoll 4U 1820--30:\\ first detection of hard X-ray emission}

\author{A. Tarana,\altaffilmark{1,2} A. Bazzano,\altaffilmark{1} P. Ubertini\altaffilmark{1} and A. A. Zdziarski\altaffilmark{3}}
\altaffiltext{1}{Istituto di Astrofisica Spaziale e Fisica Cosmica-INAF,
via del Fosso del Cavaliere 100, I-00133 Roma}\email{antonella.tarana@iasf-roma.inaf.it}
\altaffiltext{2}{Dipartimento di Fisica, Universit\`a di Roma Tor Vergata, via della Ricerca Scientifica 1, I-00133 Roma}
\altaffiltext{3}{Centrum Astronomiczne im.\ M. Kopernika, Bartycka 18, 00-716 Warszawa, Poland}
\altaffiltext{$^\star$}{INTEGRAL is an ESA project with instruments and Science Data Centre funded by ESA member states (especially the PI countries: Denmark, France, Germany, Italy, Switzerland, Spain), Czech Republic and Poland, and with the participation of Russia and the USA.}

\begin{abstract}
We study the 4--200 keV spectral and temporal behaviour of the low mass X-ray binary \source\ with {\it INTEGRAL\/} during 2003--2005. This source as been observed in both the soft (banana) and hard (island) spectral states. A high energy tail, above 50 keV, in the hard state has been observed for the first time. This places the source in the category of X-ray bursters showing high-energy emission. The tail can be modeled as a soft power law component, with the photon index of $\simeq$2.4, on top of thermal Comptonization emission from a plasma with the electron temperature of $kT_{\rm e}\simeq 6$ keV and optical depth of $\tau\simeq 4$. Alternatively, but at a lower goodness of the fit, the hard-state broad band spectrum can be accounted for by emission from a hybrid, thermal-nonthermal, plasma. During this monitoring the source spent most of the time in the soft state, usual for this source, and the $\ga$4 keV spectra are represented by thermal Comptonization with $kT_{\rm e}\simeq 3$ keV and $\tau \simeq 6$--7.
\end{abstract}

\section{Introduction}

4U 1820--30 is a low mass X-ray binary seen at 0.66$^{\prime\prime}$ from the centre of the globular cluster NGC 6624. It was the first identified source of type-I X-ray bursts \citep{grindlay}. Distance has been estimated as $d\simeq 5.8$--8.0 kpc (e.g., Shaposhikov \& Titarchuk 2004; Kuulkers et al.\ 2003). Kuulkers et al.\ (2003) found their determination of $d=7.6\pm 0.4$ kpc compatible with the peak X-ray burst luminosity being a standard candle in their sample of sources, and equal to the Eddington limit for He. The binary system consists of a He white dwarf with mass of 0.06--$0.08 M_{\sun}$ \citep{rappa} and a neutron star, with the mass estimated by Shaposhnikov \& Titarchuk as $\sim\! 1.3M_{\sun}$, orbiting at short period of 11.4 minutes \citep{Stella}. In X-rays, \source\ is classified as an atoll \citep{hasinger}. However, its flux variation between the soft (banana) to the hard (island) state are quasiperiodic at $\sim$170 d (Priedhorsky \& Terrell 1984; {\v S}imon 2003; Wen et al.\ 2006), which quasiperiodicity has been proposed to be due to tidal effects of a more remote third star (Chou \& Grindlay 2001; Zdziarski, Wen \& Gierli\'nski 2006). X-ray bursts occur only at low flux levels (Cornelisse et al.\ 2003) and the kHz QPOs are correlated with the flux (Zhang et al.\ 1998), proving that the variability is due to an intrinsic change of the source luminosity (and the accretion rate) rather than being due to variable either obscuration or viewing angle of an anisotropic emitter. X-ray spectra of the source from {\it RXTE\/} and {\it BeppoSAX\/} were fitted by thermal Comptonization with or without an additional blackbody (Bloser et al.\ 2000; Sidoli et al.\ 2001; Migliari et al.\ 2004).

In this paper, we report the first detection of X-ray emission above 50 keV from \source\ in a hard spectral state. We studied the source in the 4--200 keV energy range with the JEM-X \citep{lund} and IBIS \citep{uber} instruments on board the {\it INTEGRAL\/} satellite \citep{wink}, collected during the Galactic Centre Deep Exposure (GCDE) program \citep{wink2}. We combined three methods of analysis: {\it Temporal}, using the light curves in different energy bands; {\it Photometric}, with the Hardness--Instensity Diagram; and {\it Spectral}, by spectral fitting.

\section{Observations and Data Analysis}

The coded mask imager IBIS on board {\it INTEGRAL\/} has a field of view (FOV) of $19 \degr \times$19$\degr$ or $9\degr\times 9\degr$ when partially or fully coded, respectively. The X-ray monitor JEM-X (4--35 keV) has the fully coded FOV of $4.8\degr\times 4.8\degr$. An {\it INTEGRAL} revolution lasts 3 days, and consists of adjacent pointings of $\sim$2000 s, called Science Windows (SCW; Winkler et al.\ 1999).

We have analysed all the available data during which the source was within the IBIS/ISGRI and JEM-X fully coded FOV, so that the flux evaluation is not affected  by calibration uncertainties in the off-axis response. This yields 308 and 51 SCWs for IBIS and JEM-X respectively, during revolutions 50--363. Data were extracted with the Off-Line Scientific Analysis (OSA) \citep{gold} v.\ 5.1 software released by the {\it INTEGRAL\/} Science Data Centre (Courvoisier et al.\ 2003). A 2\% systematic errors to both JEM-X and ISGRI data sets has been added\footnote{http://isdc.unige.ch/?Support+documents}. Fluxes have been normalized to ISGRI value. The spectral analysis was performed with the XSPEC package v.\ 11.3. 

\section{Light curves and hardness--intensity Diagrams}

\begin{figure}[b]
\centering
\includegraphics[height=11cm,angle=90]{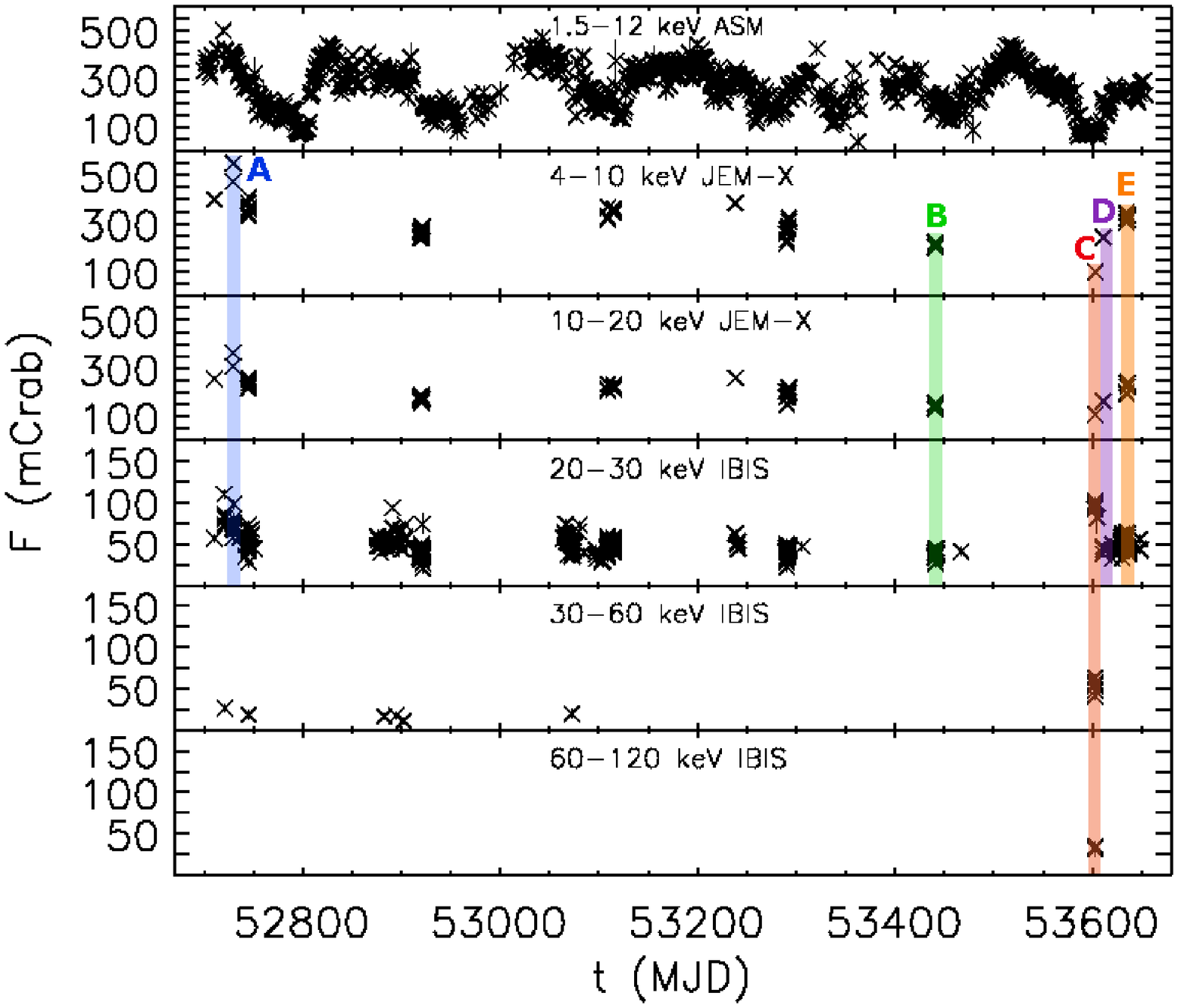}
\caption{The 2003--2005 light curves of \source, with the detector count rate in each band given with respect to the corresponding Crab count rate. The panels are marked with the energy range and detector, and present the {\it INTEGRAL\/} SCW data points except for the top panel, which gives the {\it RXTE}/ASM 1-day averages. The lines mark the data sets (A, B, C, D, E) used for joint IBIS and JEM-X spectral fits (see Table \ref{tab_dataSET}). \emph{[See on-line edition for the coloured version of this figure]}. \label{plotone}}
\end{figure}

Fig.\ \ref{plotone} shows the light curves of the monitoring period (2003 March 12--2005 October 5) from soft to hard X-rays, in the 4--10, 10--20 keV bands from JEM-X and 20--30, 30--60, 60--120 keV with IBIS/ISGRI, as well as the corresponding {\it RXTE}/ASM\footnote{http://xte.mit.edu/ASM$\_$lc.html.} light curve. The $\sim$170-d quasiperiodic variability can be seen in the ASM data.
The {\it INTEGRAL\/} light curves are marked with vertical lines to identify five epochs (A, B, C, D, E) during which we have performed detailed spectral analysis (see Table \ref{tab_dataSET}). Epoch A corresponds to the joint IBIS and JEM-X count rate maximum ($\sim$530 mCrab at 4--10 keV). Epoch C corresponds to the minimum of the 4--10 keV count rate ($\sim$100 mCrab), but to high count rates from 20 to 60 keV, implying a spectral hardening across the 4--60 keV band. We also studied epochs before (B), and after that event (D, E).

\begin{figure}[ht]
\centering
\includegraphics[height=6cm,angle=90]{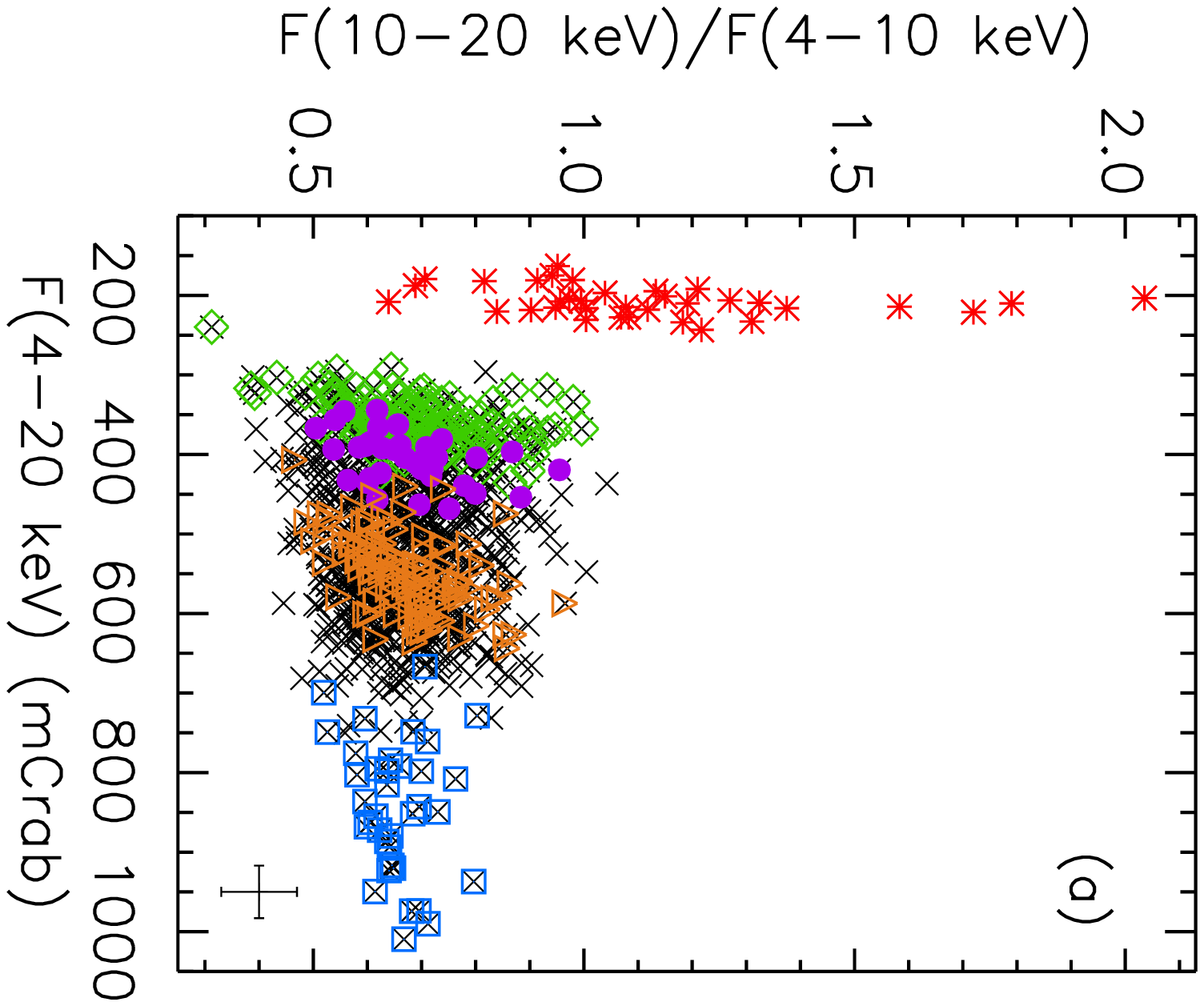}
\includegraphics[height=6cm,angle=90]{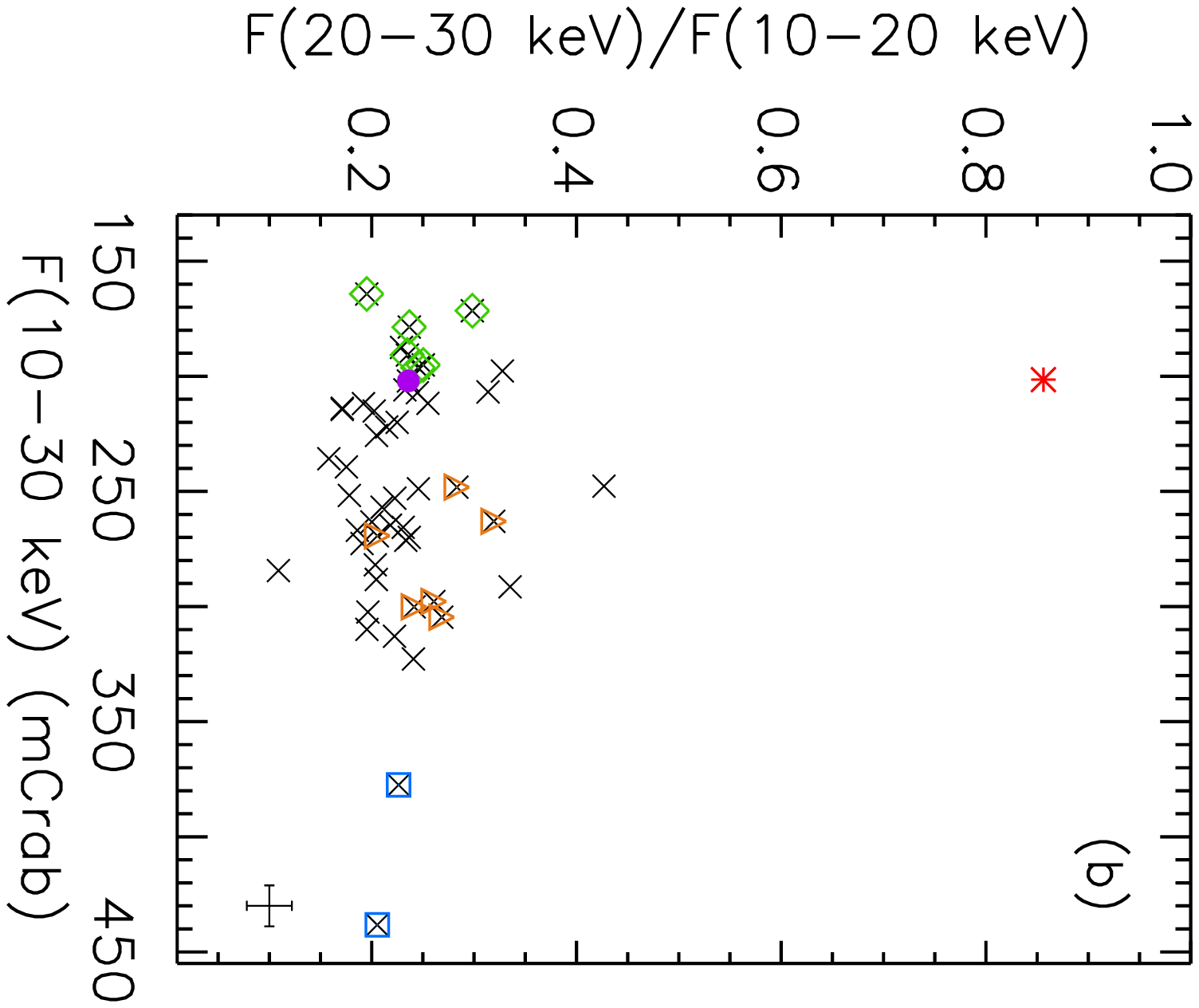}
\caption{(a) Hard colour--intensity diagram; each point corresponds to 100 s. (b) Very hard colour--intensity diagram; each point corresponds to a SCW. The colours identify the A--E data sets, see Table \ref{tab_dataSET} and Fig.\ \ref{plotone}. The typical error bars are shown at the bottom right corners.
\label{plottwo}}
\end{figure}

Fig.\ \ref{plottwo} shows hardness--intensity diagrams with data from JEM-X (a) and IBIS/JEM-X (b) (for simultaneous pointing and when possible because of different FOV). In Figs.\ \ref{plottwo} we can see the banana state, forming nearly horizontal bands across the A, B, D, E epochs. Thus, the source evolution is mostly in the flux, with the 4--20 keV spectral shape nearly constant. The upper part of the red points (epoch C) in Fig.\ \ref{plottwo}a corresponds to the island state. In this case the 4--20 keV flux is almost constant while the hardness shows large changes. This indicates spectral pivoting somewhere in this energy band. Fig.\ \ref{plottwo}b allows us to study spectral variability of the source at the energy band 10--30 keV, grater than that usually studied. In this diagram, we see similar behaviour in the horizontal banana state, while there is only one SCW for the island state, with a large hard colour. These findings are confirmed by spectral analysis as discussed below. 

\section{Spectral analysis}
\label{spectral}

\begin{figure}[b]
\centering
\includegraphics[height=8cm,angle=-90]{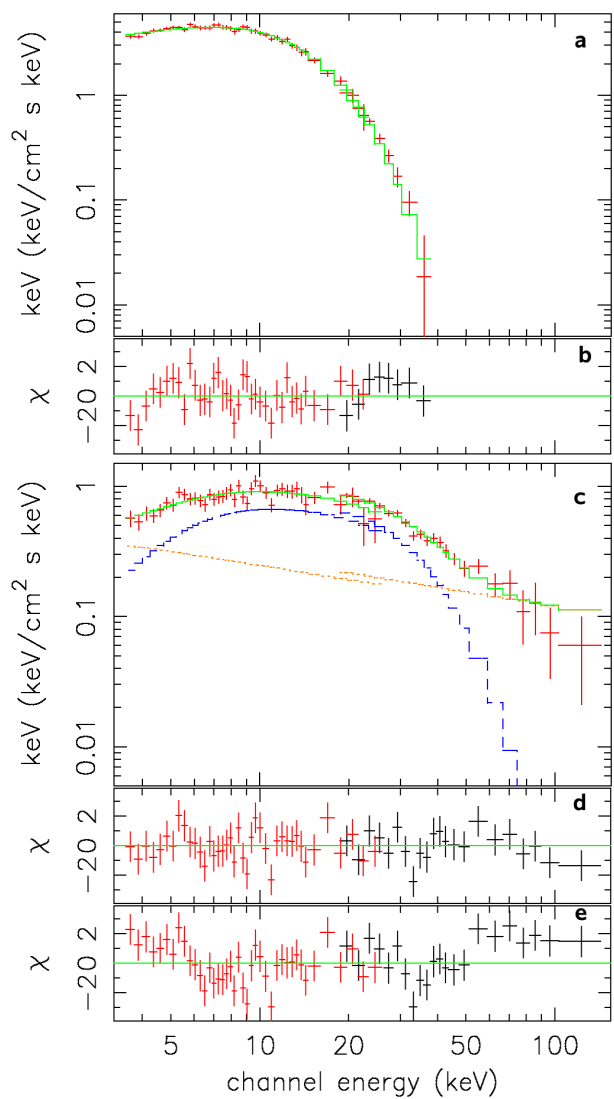}
\caption{Data, models and residuals of the soft spectrum A (\emph{a}, \emph{b}) and the hard spectrum C (\emph{c}, \emph{d}) fitted by thermal Comptonization without and with, respectively, a power law component. The panel \emph{e} shows the residuals for the spectrum C fitted without the power law component. The high energy excess at $\ga 50$ keV is clearly visible.
\label{plotthree}}
\end{figure}

Data of the five epochs (see Table \ref{tab_dataSET}) have been fitted with a number of models and their combinations, namely, thermal and hybrid Comptonization, blackbody, disk blackbody and a power law, all absorbed by the interstellar medium at $N_{\rm H}=3\times 10^{21}$ cm$^{-2}$ (Bloser et al.\ 2000). First, a thermal Comptonization model, CompTT (Titarchuk 1994), sufficiently model the spectra of the banana states (A, B, D, E). We obtain the electron temperature, $kT_{\rm e}\simeq 2.7$--3.1 keV, and optical depth $\tau\simeq 6$--7, as reported in Table \ref{tabspetot}. This is consistent with the colour-intensity diagram, where we found the spectral changes mostly in flux (most likely proportional to the accretion rate) and only slightly in their shape. The temperatures of the seed photons (Table \ref{tabspetot}) have been frozen at their best-fit values, given that those photons are entirely below the fitted band. The data, model, and residuals for the spectrum A (characterized by highest flux and unabsorbed bolometric luminosity of $7.7\times 10^{37}(d/5.8\,{\rm kpc})^2$ erg s$^{-1}$) are shown in Fig.\ \ref{plotthree}a, b, and all unabsorbed models, in Fig.\ \ref{plotfour}. An addition of a blackbody component does not significantly improve the fits (although its presence is hinted by the negative residuals at the lowest energies, Fig.\ \ref{plotthree}b), and thus the confidence ranges of the temperature, $kT_{\rm bb}$, are not constrained. Still, we find $kT_{\rm bb}\simeq 1.5$--2.4 keV at the best fits, in agreement with previous study of Bloser et al.\ (2000). 

The spectrum of the island state (C, red) shows an evident excess above 50 keV that cannot be fitted by thermal Comptonization only, as shown in Fig.\ \ref{plotthree}e (the resulted $\chi^2_\nu$ is 1.6 with 59 d.o.f.). An addition of a soft ($\Gamma\simeq 2.4$) power law improves the fit significantly ($\chi^2_\nu=1.0$, see Figs.\ \ref{plotthree}c, d), with the chance probability from the F-test (Bevington \& Robinson 1992) of $6.3\times 10^{-7}$. The parameters of the thermal plasma are also different from those for the banana state, with a higher $kT_{\rm e}$ ($\simeq 6$ keV), and a lower $\tau$ ($\simeq 4$), see Table \ref{tabspetot}. Fig.\ \ref{plotfour} shows the unabsorbed spectrum with its two components. By adding a disk blackbody component (Mitsuda et al.\ 1984) to the thermal Compton model, we have also obtained a good fit, with $\chi_\nu^2 =1.0$. However, the required inner temperature, $kT_{\rm in}\simeq 5.4$ keV, is clearly unphysical. Whereas adding a single blackbody gave a worse fit, as well as an unphysically high temperature, $\simeq 4$ keV.

We have investigated whether the tail can be an artefact of calibration inaccuracies. Analysing the data with previous software version of the OSA, the statistical significance of tail remains unchanged.

The island-state data (C) have been fitted also with the highly-accurate Comptonization model COMPPS (Poutanen \& Svensson 1996). Spectra agree very well with results from Monte Carlo Comptonization (Zdziarski et al.\ 2000, 2003). We assumed here spherical geometry of the Comptonizing plasma. Also, we have taken into account Compton reflection (Magdziarz \& Zdziarski 1995) detected in other atolls (e.g., Zdziarski, Lubi\'nski \& Smith 1999; Barret et al.\ 2000; Gierli{\'n}ski \& Done 2002), as well as a disk blackbody. We have obtained a very good fit, $\chi^2_\nu$(d.o.f.)$=0.92(54)$ for a purely-thermal Comptonizing plasma with $kT_{\rm e}\simeq 55$ keV and $\tau\simeq 0.6$, and the solid angle subtended by reflector of $\Omega/2\pi\simeq 0.69$ (assuming the source inclination of $i=60\degr$). The bolometric luminosity of this model is $1.8\times 10^{37}(d/5.8\,{\rm kpc})^2$ erg s$^{-1}$. However, similar to the problem with the model using CompTT (see above), the inner disk temperature is rather high, $kT_{\rm in}\simeq 2.4$ keV, as well as the temperature of the seed blackbody photons upscattered in the Comptonizing plasma is unphysically high, $kT_0\simeq 3.9$ keV. 

In order to constrain the model to plausible values of the blackbody temperatures, we have considered a model with $kT_0$ allowed to be $\leq 2$ keV. Also, we have constrained $\Omega/2\pi\leq 1$. The resulting purely-thermal model provides a worse fit, $\chi^2_\nu$(d.o.f.)$=1.09(54)$, with $kT_{\rm e}\simeq 24$ keV, $\tau\simeq 1.9$, $kT_{\rm in}\simeq 1.4$ keV, similar to the parameters found in the atoll source 4U 1608--52 (Gierli{\'n}ski \& Done 2002). To improve the fit of this model, in particular to account for the high-energy residuals (appearing as a high-energy tail), we have added nonthermal electrons to the thermal plasma, as done to account for some spectra of both black-hole (Gierli\'nski et al.\ 1999; Wardzi\'nski et al.\ 2002) and neutron-star (Farinelli et al.\ 2005) binaries. We have obtained $\chi_\nu^2$(d.o.f.)$ =1.07(52)$ for a model with the nonthermal tail with the power-law index of $p\simeq 1$ above the electron Lorentz factor of 1.6 (and the Maxwellian distribution below it), and other parameters similar to that of the previous model. We stress, however, that this model is significantly more complex than the one with CompTT and a power law presented above, and still yields significant residuals at energies $\ga 80$ keV. 

\section{Discussion and Conclusions}

We have presented the first detection of hard emission above 50 keV in the island state of the atoll source \source\ using IBIS. Up to now, \source\ has not been included in the sample of the high-energy emitting bursters \citep{bazzano} and indeed neither {\it BeppoSAX}, {\it CGRO}/BATSE  nor {\it RXTE\/} detected this source above 50 keV (Bloser et al.\ 2000 and references therein). The previous lack of detection might have been due to either the source spending most of the time in the soft state, the weakness of the emission above 50 keV in the island state, or the strong emission at $\ga 50$ keV appearing in that state only occasionally. 

In our island-state observation, the relative flux contribution for energies $>60$ keV is $\sim$10\%. The ratio of the 4--20 keV flux to the 20--120 keV ones is  $\sim$2, whereas it is in the range 20--27 for the soft (banana) state. We point out that we detected emission above 50 keV when the 1--20 keV luminosity was $\simeq 1.4 \times 10^{37}(d/5.8\,{\rm kpc})^2$ erg s$^{-1}$, i.e., close or somewhat below the critical value at which the X-ray binaries become hard emitters as proposed by Barret et al.\ (1996, 2000). 


While the origin of the thermal Comptonization component in the soft state appears well understood as emission from hot electrons that upscatter soft photons coming from the accretion disk and the neutron star surface, the origin of the high-energy tail is less clear.

\begin{figure}[ht]
\centering
\includegraphics[height=10cm,angle=-90]{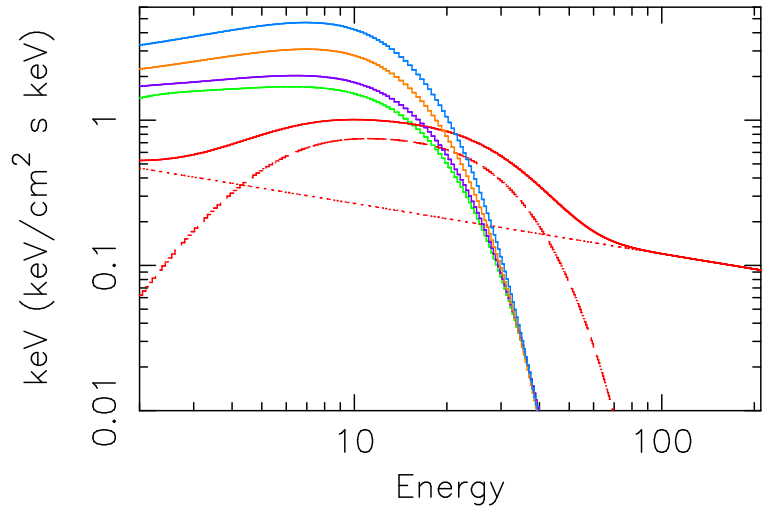}
\caption{The unabsorbed best-fit models of the A--E data sets, see Table \ref{tab_dataSET}. The model for the green, purple, orange and blue spectra is thermal Comptonization, while a power law (dotted curve) is added in the red spectrum (C). Note that the power law dominates at low energies, and thus it is probably only a phenomelogical, but unphysical, description of the spectrum. 
\label{plotfour}}
\end{figure}

The radio emission in the X-ray binaries is suggested to originate from jet-like outflows which may also produce X-rays \citep{fender04}. In particular, there are indication that both black-hole candidate and neutron stars sytems, when in hard state, produce steady jet (Migliari \& Fender 2006). Also, in the hard state, positive correlation between radio and X-ray fluxes has been reported (Migliari \& Fender 2006).
\source\ is one of the few radio detected atoll source. The only radio emission reported so far was during the soft state \citep{migliari}, but it is probable that radio emission exist also in the hard state, as e.g. for 4U 1728-34 \citep{mi03}.
Furthermore, the power-law tail detected by us is relatively soft, with $\Gamma\simeq 2.4$, and dominates over the Comptonization emission already for $E\la 4$ keV, as shown in Fig.\ \ref{plotfour}. Integration to its minimum energy, $E_{\rm min}$, yields the luminosity in this component $\propto E_{\rm min}^{2-\Gamma}$. If this component is due to nonthermal synchrotron emission, $E_{\rm min}$ corresponds to the self-absorption turnover, typically in the IR or optical range, as it is the case in another ultracompact binary, 4U 0614+091 (Migliari et al.\ 2006). Then the tail luminosity would be $\sim$10 times that of X-rays (see also discussion in Zdziarski et al.\ 2003), making the island state more luminous than the banana state, with the maximum of the overall $EF_E$ at $\sim$1 eV, neither of those seen in X-ray binaries. 

Also, the presence of the jets or outflow could be due to magnetic reconnection avents above the accretion disk (Shibata 2005).

A more likely origin of the power law tail appears to be emission of nonthermal electrons in a hybrid (i.e., containing also thermal electrons) plasma, as discussed in Section \ref{spectral}. Nonthermal tails have been commonly seen in soft states of black-hole binaries (e.g., Cyg X-1, Gierli\'nski et al.\ 1999) and neutron-star Z-sources, e.g., GX~17+2 (Di Salvo et al.\ 2000; Farinelli et al.\ 2005), Sco~X-1 \citep{damico}, GX~349+2 \citep{disalvo_b}, GX~5--1 \citep{asai}, while they are difficult to detect in the atoll sources, e.g., 4U 0614+091 \citep{pira}. On the other hand, nonthermal tails have been found in the hard state of some black-hole binaries, e.g. Cyg X-1 (McConnell et al.\ 2002) and GX 339--4 (Wardzi\'nski et al. 2002).

The \source\ high-energy part of the spectrum (E $\ga 10$ keV) can be well fitted with purely thermal Comptonization at $kT_{\rm e}\simeq 50$ keV, as discussed in Section \ref{spectral}. This electron temperature is higher than that seen in 4U 1608--52 (Gierli\'nski \& Done 2002), but the Comptonization temperature of atolls in the island state has not yet been determined for a sample of sources, and $kT_{\rm e}\simeq 50$ keV is similar to that seen in the hard state of black-hole binaries (e.g., Zdziarski \& Gierli\'nski 2004). Then, the problem with fitting the spectrum at $<10$ keV with this model may be due to the complexity of the X-ray spectra in the hard state, including contributions from the disk, the corona above the disk, optically-thin and optically-thick parts of the boundary layer and the uncovered part of the neutron star surface.

A different thermal Comptonization model has been considered by Markoff et al.\ (2005). They proposed thermal synchrotron self-Compton emission in a jet as a model for the hard state of black hole binaries. However, their model requires an extremely high electron temperature, $kT_{\rm e}\sim 3$--4 MeV. This would result in the thermal spectral cutoff in black hole binaries at energies $\ga 3kT_{\rm e}\sim 10$ MeV, which has not been seen in {\it any\/} object yet. Although we cannot formally rule out a contribution from this process, postulating the presence of $\sim$3 MeV plasma in 4U 1820--30 appears rather speculative at this time. 

In summary, we have discovered strong high-energy emission from 4U 1820--30. The high energy emission is harder than that from some other atolls in the island state. Its most likely interpretation is either emission of some nonthermal electrons or thermal emission from plasmas with a relatively high temperature, $kT_{\rm e}\ga 50$ keV. 

\acknowledgments
This research has made use of data obtained through the {\it INTEGRAL\/} Science Data Centre, Versoix, Switzerland. The authors thanks M. Federici for the continuous effort to update the {\it INTEGRAL\/} archive and software in Rome and G. De Cesare, L. Natalucci and F. Capitanio for the scientific and data analysis support. Particular thanks are due to P. Goldoni for making his private data available for analysis. This work has been supported by the Italian Space Agency through the grant I/R/046/04. AAZ has been supported by the Polish grants 1P03D01827, 1P03D01128 and 4T12E04727.

\clearpage

%
\begin{table}[t]
\begin{center}
\caption{The log of the data used for spectral fitting. The indicated colours are used in the light curves, Fig.\ \ref{plotone}, and hardness--intensity plots, Fig.\ \ref{plottwo}.
\label{tab_dataSET}}
\begin{tabular}{lcccc}
\tableline
data set & Time (MJD) &  rev. \#  & Exposure (ks) & Obs.\ type\\ 
& start$-$stop & & IBIS ; JEM-X\tablenotemark{a} & \\
\tableline
A (blue) &52729.2$-$52729.5  & 56  & 6.7 ; 3.7 &GCDE     \\
B (green)&53440.6$-$53442.3  & 294  &1.9 ; 1.3  &  GCDE      \\
C (red)   &53602.4$-$53603.3 & 348  & 1.8 ; 3.5 & ToO\tablenotemark{b}     \\
D (purple) &53610.9$-$53620.2 & 351  & 1.2 ; 3.5 & ToO\tablenotemark{b}, GCDE\\
E (orange) &53634.8$-$53635.5 & 359 &  1.3 ; 1.2  &  GCDE   \\
\tableline
\end{tabular}
\vspace{-0.4cm}
\tablenotetext{a}{JEM-X2 for the A obs. and JEM-X1 for the B-E obs.}
\tablenotetext{b}{AO3 target-of-opportunity observation, PI P. Goldoni.} 
\end{center}
\end{table}

\begin{table*}[ht]
\begin{center}
\caption{Spectral fitting results for the JEM-X and IBIS broad-band spectra. The model is CompTT for A, B, D, E, and a CompTT + a power law for C. 
\label{tabspetot}}
\begin{tabular}{l|ccccc}
\tableline
parameters & A (blue) & B (green)  &  C (red)   &  D (purple) & E (orange)\\
\tableline
$kT_{0}$ (keV)\tablenotemark {a}         & 0.2                  & 0.4                    & 1.5                    & 0.3                    & 0.2 \\
$kT_{\rm e}$ (keV) &2.71$^{+0.08}_{-0.07}$ & 3.12$_{-0.10}^{+0.10}$ & 6.10$^{+0.77}_{-0.62}$ & 2.90$_{-0.18}^{+0.20}$ & 2.83$_{-0.06}^{+0.07}$\\
$\tau$                         &7.04$^{+0.32}_{-0.29}$ & 5.85$_{-0.23}^{+0.24}$ & 3.89$_{-0.65}^{+0.79}$ & 6.38$_{-0.62}^{+0.72}$ &6.71$_{-0.22}^{+0.24}$ \\
norm$_{\rm CompTT}$                &2.34$^{+0.21}_{-0.19}$ & 0.58$^{+0.06}_{-0.06}$ & 3.79$^{+0.68}_{-0.59}$ $\times 10^{-2}$ & 1.06$_{-0.18}^{+0.21}$&1.57$_{-0.11}^{+0.12}$ \\
\tableline
$\Gamma$    & $-$ &$-$  & 2.35 $_{-0.11}^{+0.10}$ &$-$ & $-$  \\
norm$_{pl}$ & $-$ &$-$  & 0.60$_{-0.24}^{+0.22}$ &$-$ & $-$ \\
\tableline
$\chi_\nu^2$(d.o.f) &1.13(44) & 0.96(46)&1.0(57) & 1.07(37) & 1.14(43)\\
\tableline
$F_{\rm 4-20 keV}$ (erg s$^{-1}$ cm$^{-2}$)  &9.1$\times 10^{-9}$ & 3.8$\times 10^{-9}$ &2.1$\times 10^{-9}$  &4.2$\times 10^{-9}$  & 5.8$\times 10^{-9}$\\
$F_{\rm 20-60 keV}$ (erg s$^{-1}$ cm$^{-2}$) &3.3$\times 10^{-10}$& 1.8$\times 10^{-10}$&7.4$\times 10^{-10}$  &2.2$\times 10^{-10}$ & 2.5$\times 10^{-10}$  \\
$F_{\rm 60-120 keV}$ (erg s$^{-1}$ cm$^{-2}$)& $-$ & $-$ &1.5$\times 10^{-10}$ & $-$ & $-$ \\
\tableline
\end{tabular}
\vspace{-0.4cm}
\tablenotetext{a}{ Fixed parameters}
\end{center}
\end{table*}

\end{document}